\newcommand{\CLASSINPUTtoptextmargin}{0.76in}
\newcommand{\CLASSINPUTbottomtextmargin}{1.03in}
\documentclass[conference]{IEEEtran}
\IEEEoverridecommandlockouts
\usepackage{cite}
\usepackage{amsmath,amsfonts,amsthm,amssymb,amsbsy,amstext,amscd,amsxtra,multicol,enumitem,soul}
\usepackage{mathtools, stackrel,multirow}
\usepackage{algorithmic}
\usepackage{graphicx}
\usepackage{textcomp}
\usepackage{xcolor}
\usepackage{tikz}
\usepackage{textcomp}
\usepackage{hyperref}
\usepackage{stmaryrd,scalerel} 

\def\BibTeX{{\rm B\kern-.05em{\sc i\kern-.025em b}\kern-.08em
    T\kern-.1667em\lower.7ex\hbox{E}\kern-.125emX}}

\def\l{\mathbf l}
\def\N{\mathbb N}
\def\R{\mathbb R}

\let\lvec\overleftarrow
\let\rvec\overrightarrow
\let\lrvec\overleftrightarrow 
\def\div{\mathrm{div}}

\makeatletter
\newcommand{\linebreakand}{%
  \end{@IEEEauthorhalign}
  \hfill\mbox{}\par
  \mbox{}\hfill\begin{@IEEEauthorhalign}
}
\makeatother

\newcommand\copyrighttext{%
  \footnotesize \textcopyright 2012 IEEE. Personal use of this material is permitted.
  Permission from IEEE must be obtained for all other uses, in any current or future 
  media, including reprinting/republishing this material for advertising or promotional 
  purposes, creating new collective works, for resale or redistribution to servers or 
  lists, or reuse of any copyrighted component of this work in other works. 
  DOI: \href{<http://tex.stackexchange.com>}{<DOI No.>}}
\newcommand\copyrightnotice{%
\begin{tikzpicture}[remember picture,overlay]
\node[anchor=south,yshift=10pt] at (current page.south) {\fbox{\parbox{\dimexpr\textwidth-\fboxsep-\fboxrule\relax}{\copyrighttext}}};
\end{tikzpicture}%
}

\begin{document}

\title{Efficient Mixed Integer Linear Programming Approaches to Dynamic Path Restoration\\
{
\thanks{\textsuperscript{*} This paper was prepared within the framework of the HSE University Basic Research Program. 
}
}

\author{\IEEEauthorblockN{Alexander Rubtsov\textsuperscript{*}}
\IEEEauthorblockA{\textit{faculty of computer science} \\
\textit{HSE University}\\
Moscow, Russia \\
ORCID 0000-0001-8850-9749}
\and
\IEEEauthorblockN{Bruno Bauwens\textsuperscript{*}}
\IEEEauthorblockA{\textit{faculty of computer science} \\
\textit{HSE University}\\
Moscow, Russia \\
ORCID 0000-0002-6138-0591\\
brbauwens@gmail.com}
\and
\IEEEauthorblockN{Dmitri Shmelkin}
\IEEEauthorblockA{\textit{Independent University of Moscow,} \\
Moscow, Russia \\
ORCID 0000-0001-9815-6115\\
Dmitri.Shmelkin@gmail.com}
\linebreakand  
\IEEEauthorblockN{Elizaveta Rudenko}
\IEEEauthorblockA{\textit{faculty of mathematics} \\
\textit{HSE University}\\
Moscow, Russia \\
elizaveta.4030@gmail.com}
\and
\IEEEauthorblockN{Alexey Lavrov}
\IEEEauthorblockA{\textit{faculty of mathematics} \\
\textit{HSE University}\\
Moscow, Russia \\
xnikiv@gmail.com}
}
}

\maketitle

\begin{abstract}
  We consider the problem of single link failure in an elastic optical network, (also known as flex-grid WDM network).
  The task is to reroute optical connections that go through the broken link using free capacity of other links of the network. 
  Nowadays, dynamic restoration gains popularity, in which the possiblity of rerouting is only inspected after a link failure is detected. 
  Since the problem of recovery is NP-hard, heuristic algorithms are used to either find such routes, or suggest that the routes do not exist. 
  In order to understand the quality of these heuristics, often mixed integer linear programming is used to obtain exact positive and negative answers. 
  We present a detailed such model that checks whether restoration is possible without the use of additional regenerators. 
  This means, that the new light paths need to satisfy a length constraint. 
  As preprossing we apply a trimming procedure that takes advantage of this length constraint, and significantly speeds up the evaluation of these models. 
  Our model is more general, and besides solving the problem of link restoration, also solves the full problem of wavelength and spectrum assignment. 
\end{abstract}


\section{Introduction}

\copyrightnotice
Nowadays, optical communication networks may contain up to thousands of links, and in such networks, some optical wire gets failure regularly due to various reasons. Because network sizes are increasing, this happens more frequently, and additionally, link failure may impact services with ingress/egress nodes at a larger distance from the failed link.
Typically, fixing such a failure may take hours, because some operator staff needs to go to the place of the cut. 
In the mean time, the network traffic through the broken link needs to be rerouted, and preferably this is done within a second or faster. 

Such link restoration is possible because some network resources remain vacant (on purpose). Every optical link provides multiple frequencies using wavelength division and multiplexing (WDM for short) technology. For a general introduction to the technology, we refer to~\cite{Simmons2014optical}. 
A client orders a connection between 2 nodes of the network, and these nodes are connected by a light path of a single color (for long connections, sometimes, several unicolored segments are used). 
Typically, in a link, there is some fraction of free colors, and these colors may be used for rerouting the broken traffic after link failure. 
A traditional approach to service resilience is widely used such that some services are provided with pre-computed secondary path(s), and these will be enabled if some link of the primary path fails, see the books~\cite{Grover2003mesh, Vasseur2004network, Mukherjee2005survivable} for background. 
In~\cite{popularRestorationI}, the details are given of a mixed integer linear programming approach for several such protection and restoration strategies.\footnote{
  This approach is connected to a specific group of wavelength and spectrum assignment strategies, called `alternate routing', 
  where for each request, only a fixed set of paths can be used, and this set is used in the MILP model. 
  } The disadvantage of the traditional approach is low level of resource re-use so that such an approach is rather resource inefficient. 
For background on (integer) linear programming, we refer to~\cite{Zang2000Review, Oki2013linear}.

To cope with the above disadvantage, dynamic restoration is used, which is a method that only considers restoration after it is known which link is broken. The challenges of this approach are slower path deployment and possible failures to recover connections of all clients. 
On the other hand, this approach allows deploying more services and scales better for large networks, see a rather old comparison of these approaches in ~\cite{Doucette2001comparison}.

A second recent trend is that demands may have varying capacities, typically called {\em widths}. In the older fix-grid setting, all these widths are the same, but in the flex-grid setting, they can be different. These flex-grid networks are also called {\em optical elastic networks}, and we refer to the book~\cite{Chatterjee2020elastic} for background. In the paper \cite{Paolucci2014multipath}, you may find heuristic algorithms for restoration in this setting. 

\bigskip
We now proceed to the detailed formulation of the problem. 
There is a network of devices connected via optical fiber links. 
The signal transmitted from a sender node $s$ to a receiver node $t$ occupies several consecutive frequencies, 
so the frequency spectrum is divided into $C$ frequency slots, where $C$ is a constant which is the same for each link. 
The number of slots~$w$, the {\em width} is fixed for each signal and depends on technical needs. 
Moreover, the signal goes through several fiber links adjacent to intermediate nodes.
Each link has a length (in kilometers) and for robust transmission of the signal, 
the length of the chosen path from $s$ to $t$, i.e., the sum of fiber links length's, should not exceed some reachability threshold~$r_d$, which may depend on the demand because of the optical receiver properties.
A \emph{demand} is a tuple $d = (s_d, t_d, w_d, r_d)$, and each demand is associated to a signal (in the busy network). 
A {\em path} is a tuple $P = (\mathbf{l}, c, w)$, where $\l = l_1, \ldots, l_m$ is a sequence of adjacent links, where $c \in \{1, \ldots, C\}$ is the first occupied slot, and $w$ is the {\em width}, which is the number of slots. Thus the path occupies slots $c, c+1,\ldots, c+w-1$ of each fiber link $l_i \in \l$. A path $P$ is a \emph{valid path} for a demand $d$, if $\l$ connects $s$ with $t$ and the length of $P$ is at most $r$. We say that paths $P$ and $P'$ do not intersect if they do not occupy the same slot of the same link. 
In our further construction, slots corresponds to colored edges of a graph, so we refer to slots as \emph{colors}.

Initially, there is an optical network with demands $d_i$ and corresponding valid non-intersecting paths $P_i$ which are used to route the signals corresponding to the demands. Because of some accident, one fiber link $l_b$ was broken, and all the paths that contain $l_b$ cannot be used for the signal transmission anymore, so we call them \emph{broken}. We denote by $D = \{d_1, \ldots, d_k\}$ the set of \emph{broken demands} corresponding to the broken paths.
The problem is to find non-intersecting valid paths for all broken demands and if this is not possible, declare a maximum subset (by the number of demands) $D' \subseteq D$ such that all demands from $D'$ can be rerouted. We call the first problem \emph{a restoration problem} and the second problem \emph{a maximum subset problem}. 

In fact, we address a more general problem than the described recovering problem that comes from practice. 
We consider an optical network with partially occupied slots. Take on the input the set of demands $D = \{d_1, \ldots, d_k\}$ and verify whether it is possible to route all demands from $D$, i.e., find non-intersecting valid paths $P_i$'s corresponding to $d_i$'s, or prove that it is impossible. And we solve the maximum subset problem for the generalized problem as well. The generalized problem is a generalization of a well-known Routing and Spectrum Allocation Problem (RSA). The original RSA problem is the case of the generalized problem, where all slots in the network are free. So we call our problem a \emph{generalized RSA} problem.

The RSA problem is known to be NP-complete (because the edge disjoint paths problem is NP-complete even for planar graphs \cite{Middendorf1993}, and this can be viewed as a special case of path restoration). In practice, an instance of a restoration problem should be solved in at most a second, so various heuristics and approaches from the combinatorial optimization are used to achieve this bounds on practical test cases. To investigate the hard cases to these approaches it is also important to have a tool that computes the \emph{ground truth}~--- ~answers for the instances~---~in a reasonable time that can be used for the evaluation of algorithms, but may be slow for directly using on practice. In this paper we are focusing on the approach for ground truth computation based on Mixed Integer Linear Programming (MILP). We present MILP models for generalized RSA problem and for the maximum subset problem as well. 

While our models are based on the same ideas as known models for the RSA problem \cite{Bertero2018integer}, it addresses a more general problem and uses fewer variables. 
For example, we do not use consecutive ranges of variables corresponding to slots. 

\section{Definitions and Notation}

In this section we define the generalized RSA problem in a more formal style and introduce the notation
needed for the further exposition, particularly for the description of MILP models.

We adapt standard notation from Graph theory~\cite{Diestel2018graph}. 
Particularly, we denote by $V(G)$ and $E(G)$ the set of nodes and the multiset of edges of the graph $G$ respectively.
We consider weighted graphs with parallel edges, so an edge is a pair $(\{u, v\}, \delta)$, 
where $\{u, v\}$ is an undirected edge, i.e., the set of size 2, so $u \neq v$, 
and $\delta$ is the weight of the edge. 

\begin{itemize}
    \item $V$ is the set of nodes of the optical network.
    \item A \emph{link} (fiber link) $l$ is a tuple $(\mathrm{id}, u, v, \delta)$, where
        \begin{itemize}
            \item $u, v \in V$ are the pair of nodes connected by $l$. While the signal can be transmitted both from $u$ to $v$ and from $v$ to $u$, the order of $u$ and $v$ in the tuple is important for technical needs.
            \item $\mathrm{id} \in \N$ is the links identifier, needed since there can be parallel links that connect the same nodes $u$ and~$v$,
            \item $\delta \in \R_{\geq 0}$ is the length of the link.
        \end{itemize}
        We refer to components of a link $l$ as $(\mathrm{id_l}, u_l, v_l, \delta_l)$.
    \item $C_l \subseteq \{1,\ldots, C\}$ sets of available slots (\emph{colors}) for a link~$l$.
    \item \emph{Optical network} $N = (V, L, l \mapsto C_l)$ is a tuple of nodes~$V$, set of links~$L$, and a mapping that returns for a link~$l$ the set~$C_l$. 
    \item We denote a multiset via braces $\lbag \cdots \rbag$, the rest of notation for multisets is adopted from the classical notation for sets.
    \item $G_c = (V, \lbag (\{u_l, v_l\}, \delta_l) : [l \in L] \land [c \in C_l] \rbag)$ 
          is an undirected weighted graph on nodes~$V$ 
          with (may be parallel) edges corresponding to links with available slot~$c$.            
          We call an edge of a graph $G_c$ an \emph{edge of color} $c$ 
          and we call $G_c$ a \emph{graph of color} $c$.
    \pagebreak      
    \item An optical network $N$ has the \emph{corresponding graph}
          $$ G_N = \bigcup\limits_{c = 1}^C G_c $$
          that is a union of colored graphs. 
          Formally, vertices of $G_N$ are labeled by its colors, i.e., the endpoints of an edge
          $(\{u, v\}, \delta) \in E(G_c)$ are labeled as $(u, c)$ and $(v, c)$ in $G_N$.
          So, $V(G_N) \subseteq V\times \{0, \ldots, C\}$. 
          For simplicity, we operate further with $G_N$'s subgraphs $G_c$ instead of $G_N$ itself
          assuming they are the parts of $G_N$.
    \item We refine the definition of a demand given in Introduction. 
          A \emph{demand }$d$ is a tuple $(\mathrm{id}_d, s_d, t_d, w_d, r_d)$
          \begin{itemize}
            \item $s_d, t_d \in V$ are the demand endpoints;
            \item $w_d$ is the number of slots (colors) required by $d$;
            \item $\mathrm{id_d} \in \N$ is the demand identifier, needed since there can be with 
                  the same values of $(s_d, t_d, w_d)$;
          \end{itemize}
    \item A path $P_d$ for a demand $d$ is a pair $(\mathbf{l}, c)$, where
    \begin{itemize}
          \item $\l = l_1, \ldots, l_m$ is a sequence of links, such that
          \begin{itemize}
              \item $s_d \in \{u_{l_1}, v_{l_1}\} $, $t_d \in \{u_{l_m}, v_{l_m}\} $
              \item $|\{u_{l_i}, v_{l_i}\} \cap \{u_{l_{i+1}}, v_{l_{i+1}}\}| = 1$
              \item the path begins in the node $s_d$ and ends in the node $t_d$, i.e.
              $s_d \not\in \{u_{l_1}, v_{l_1}\} \cap \{u_{l_{2}}, v_{l_{2}}\}$ and 
              $t_d \not\in \{u_{l_{m-1}}, v_{l_{m-1}}\} \cap \{u_{l_{m}}, v_{l_{m}}\}$              
          \end{itemize}
          \item $c \in \{1, \ldots, C\}$ is a color that is interpreted as the first color occupied by the demand. 
    \end{itemize}
    \item A path $P_d = (\mathbf{l}, c)$ for the demand $d$ is \emph{valid}
          if $\delta_{l_1}+\dots+\delta_{l_m} \leq r_d$ and for each color $c' \in \{c, \ldots, c+w_d-1\}$ the graph $G_{c'}$
          contains a path $\{u_{l_1}, v_{l_1}\}, \ldots \{u_{l_m}, v_{l_m}\}$.
    \item Paths $P_d = (\mathbf{l}, c)$ and $P_{d'} = (\mathbf{l'}, c')$ are \emph{intersecting}
          if $\exists c' \in \{c, \ldots, c+w_d-1\} \cap \{c', \ldots, c'+w_{d'}-1\}$
          and sequences $\mathbf{l}$ and $\mathbf{l'}$ contain the same link.
    \item The \emph{generalized RSA problem} takes as input an optical network $N$ and a set of demands $D$.
          The problem is to decide whether there exists a set of pairwise nonintersecting paths $P_d, d \in D$, 
          such that a path $P_d$ is a valid path for the demand $d$. In the case of the positive answer, the answer
          also shall include the paths $P_d$'s.
    \item The \emph{maximum subset generalized RSA problem} has the same input 
          and asks to find a subset $D' \subset D$ of maximal size
          such that the answer for the instance $(N, D')$ is positive.
          
\end{itemize}

\section{Trimming}\label{sect:Trimming}

    We applied our approach for the generalized RSA problem to instances raised from a practically motivated restoration problem (described in Introduction). So the optical network $N$ from the input was constructed by an original network in which a link had been broken and $D$ is the set of broken demands (the paths which went through the broken link). It occurs that in many cases some colors from sets $C_l$ can not be used for any recovered path (for any broken demand) so these can be removed and this decreases the size of the problem. We call such colors \emph{useless}. Formally, a color $c\in C_l$ of the link $l$ is \emph{useless} for a demand $d\in D$ if there is no path from $s_d$ to $t_d$ of length at most $r_d$ in the graph $G_c$; we also call the triple $(d,l,c)$ useless. A color $c\in C_l$ of the link $l$ is useless if it is useless for all $d\in D$. Since our MILP model has Boolean variables $x^d_{l, c}$ corresponding to the triples $(d, l, c)$ such that $x^d_{l, c} = 1$
iff $d$ occupies color $c$ of the link $l$, it follows that
if the triple $(d,l,c)$ is useless, then $x^{d,l}_{c}$ is always set to zero, so we can get rid of this variable and decrease the model size. 
    
So in this section we provide an algorithm that finds all useful (i.e., not useless) triples $(d, l, c)$ 
and we call by \emph{trimming} the removal of useless colors from $C_l$ and the removal of always zero variables $x^d_{l, c}$ from a MILP model as well.

The algorithm finds all useful triples $(d, l, c)$ as follows.  
Fix a demand $d$. For each $c \in \{1, \ldots, C - w_d\}$, consider the graph $G_{c:w_d}$ that contains an edge $e$ if and only if each graph $G_{c}, ..., G_{c+w_d-1}$ contains $e$. Thus, 
\begin{equation}
    G_{c:w_d} = \bigcap\limits_{c' = c}^{c+w_d-1} G_{c'}.
\end{equation} 
Now, we mark all edges in this graph that lie on at least one valid path, i.e., for which at least one of the following conditions holds:
\begin{align}
 \rho(s_d, u) + \delta + \rho(t_d, v) & \leq r_d \label{eq:trim1}\\
 \rho(s_d, v) + \delta + \rho(t_d, u)  &\leq r_d.\label{eq:trim2}
\end{align}
So any of conditions~(\ref{eq:trim1}-\ref{eq:trim2}) implies that a triple $(d, l, c')$ is useful for each $c' \in \{c, \ldots, c+w_d-1\}$.
And vice versa: if a triple $(d, l, c')$ is useful, there exists a color $c$ such that $c' \in \{c, \ldots, c+w_d-1\}$ and the graph $G_{c:w_d}$ contains a path from $s_d$ to $t_d$ of length at most $r_d$ that contains the edge~$e_l$, so at least one of conditions~(\ref{eq:trim1}-\ref{eq:trim2}) holds. Therefore our algorithm marked all useful edges and all unmarked edges are useless.   

Checking conditions~(\ref{eq:trim1}-\ref{eq:trim2}) was done using a minor modification of Dijkstra's algorithm that in 1 run computes the distance from $s_d$ to all nodes $v$. Similar for $t_d$.
This was implemented using Python's heapq datastructure. (Checking conditions~(\ref{eq:trim1}-\ref{eq:trim2}) in each graph $G_{c:w_d}$ takes time $O(m \log m)$, where $m$ is the number of edgs in the graph.) In our experiments, this takes a negligable fraction of the computation time.

In some cases, trimming can also prove that RSA is unsolvable. If for some demand $d$ the length of shortest path exceeds $r_d$ for each $G_{c:w_d}$, then the generalized RSA problem is unsolvable. We call such demands \emph{non re-routable}. So we will also find all non re-routable demands during the trimming process to remove them from the set $D$ when solving the maximum subset problem.


\section{MILP model for the generalized RSA problem}

During the run of the algorithm from Section~\ref{sect:Trimming} (of finding useful edges) we also mark a color $c$ as a valid first color for a demand $d$ if the graph $G_{c:w_d}$  contains at least one path from $s_d$ to $t_d$ of length at most $r_d$.
We denote by $C_{dl} \subseteq C_l$ the set of colors that are available for the link $l$ that can be first color for the demand $d$.
We denote the set of all useful $(d, l, c)$ triples by $U$.
 Our MILP model is based on the max-flow approach, so despite the graph $G_N$ corresponding to the optical network is an undirected graph, we shall treat its undirected edges as the pair of directed edges.

Therefore we denote by $\lvec l$ a directed version of the link $l$ with the direction $u_l \to v_l$, and 
by $\rvec l$ a directed version of the link $l$ with the direction $v_l \to u_l$.
We call $\lvec l$ and $\rvec l$ \emph{directed links} for the sake of brevity.



\vspace*{0.1mm}

We also introduce the notation $\lvec L = \{\lvec l : l \in L\}$, $\rvec L = \{\rvec l : l \in L\}$, 
and $\lrvec L = \lvec L \cup \rvec L$. 
We also define a directed variant $\lrvec U$ of useful triples: 
$\lrvec U$ contains both triples $(d,\lvec l, c)$ and $(d,\rvec l, c)$ for each $(d, l, c) \in U$.


As we announced above, our model has Boolean variables 
\begin{equation}
x^{d, l}_{c} = 
\begin{cases*}
\multirow{3}{*}{1,} & if the color $c$ of $l \in \lrvec L$\\
                    & is occupied by the path \\
                    & for the demand $d$;\\
0, & otherwise.
\end{cases*}
\end{equation}

Denote by $N_{+}(u)$ the set of directed links with left endpoint $u$ and $N_{-}(u)$ the set of links with right endpoint $u$. We adopt the commonly used notion of divergency from flow-based MILP models for our special case taking into account the colors of the links:


\begin{align}
&\div_{d,c}(u) = \sum_{\substack{(d, l, c) \in \lrvec U \\ l \in N_+(u)}} x^{d,l}_{c} - \sum_{\substack{(d, l, c) \in \lrvec U \\ l \in N_-(u)}} x^{d,l}_{c} 
\end{align}

Hereinafter, by the stack of conditions in summations we mean the conjunction of the conditions. 

Now we describe the model (\ref{mixedmilp:objective}-\ref{mixedmilp:continious:spectrum3}), then explain the intuition behind it and prove the correctness. We write in the model $\forall c \in C$ instead of $\forall c \in \{1,\ldots, C\}$ for the sake of brevity.


\allowdisplaybreaks
\begin{align}
    & \sum_{ (d, l, c) \in \lrvec U} x^{d,l}_{c} \to \min &&  \label{mixedmilp:objective}\\ 
    &                   & & \forall c\in C, \notag  \\
    & \div_{d,c}(u) = 0 & & \forall d\in D  \label{mixedmilp:div}\\
    &                   & & \forall u\in V\setminus\{s_d, t_d\}  \notag \\
    & && \notag\\
    &\sum_{\substack{l \in N_+(s_d) \\ (d, l, c) \in \lrvec U}} x^{d,l}_{c} = w_d & & \forall d\in D  \label{mixedmilp:div:s}\\
    &\sum_{\substack{l \in N_-(s_d) \\ (d, l, c) \in \lrvec U}} x^{d,l}_{c} = 0 & & \forall d\in D    \label{mixedmilp:div:s0}\\
    &\sum_{ (d, l, c) \in \lrvec U} x^{d,l}_{c}\cdot\delta_l \leq r_d\cdot w_d & & \forall d\in D \label{mixedmilp:reachability}\\
    & && \notag  \\
    &\sum_{\substack{d \in D \\ (d, l, c) \in  U}}\left(x_{c}^{d, \rvec{l}} + x_{c}^{d, \lvec{l}} \right)\leq 1 
        && \begin{array}{l}\forall c\in C,\\ \forall {l} \in {L} \end{array} \label{mixedmilp:unicolor}
\end{align}
\vspace*{-1mm}
\begin{align}      
    & \sum\limits_{\Delta c = 0 }^{w_d-1} x^{d, l}_{c+\Delta c} \geq w_d\cdot\left(x^{d,l}_{c} - x_{c-1}^{d, l}\right) 
        &&  \begin{array}{l}
                \forall l \in \lrvec{L}\\
                \forall d \in D \\
                \forall c: [c \in C_{dl}, \\
                 c-1 \in C_{dl}]
            \end{array}  \label{mixedmilp:continious:spectrum}\\
   & && \notag\\
 & \sum\limits_{\Delta c = 0 }^{w_d-1} x_{c+\Delta c}^{d, l} \geq w_d\cdot x^{d, l}_{c}
     &&  \begin{array}{l}
             \forall l \in \lrvec{L}\\
             \forall d \in D \\
             \forall c: [c \in C_{dl}, \\
              c-1 \not\in C_{dl}]
          \end{array} \label{mixedmilp:continious:spectrum2}\\
  & && \notag\\
  & x_{l,c}^{d} \leq x_{l, c-1}^{d}    
      &&  \begin{array}{l}
              \forall l \in \lrvec{L}\\
              \forall d \in D \\
              \forall c: [c \not\in C_{dl}, \\
               (d,l,c)\in \lrvec U]
           \end{array} \label{mixedmilp:continious:spectrum3}
\end{align}

In Constraints~(\ref{mixedmilp:continious:spectrum}-\ref{mixedmilp:continious:spectrum3}) we mean that the conjunction of conditions in brackets $[\cdots]$ hold.

Constraint~\eqref{mixedmilp:div} guarantees the conservation of flows in the inner vertices. Constraints (\ref{mixedmilp:div:s}-\ref{mixedmilp:div:s0}) guarantee that the value of flowing out of $s_d$ equals to the width $w_d$ and the conservation of flow implies that the incoming flow in $t_d$ has value $w_d$ as well; moreover, no flow enters $s_d$ and leaves $t_d$.  

Constraint~\eqref{mixedmilp:reachability} guarantees that the path for the demand $d$ does not exceed the length $r_d$.

Constraint~\eqref{mixedmilp:unicolor} guarantees that each color of each link is occupied by at most one demand. It is needed since there are two directed links in the MILP model for each physical link, so if a color $c$ is occupied at $\rvec l$ it is unavailable for $\lvec l$. 

To explain the intuition behind the rest of constraints we shall mention the following observation. Assume that paths for all demands have been successfully routed and the variables $x^{d,l}_{c}$ set according to the routing. Fix a color $c$; if the path for the demand $d$ traverses a link $l$ and $(d, l, c-1) \in \lrvec U$, i.e. the color $c-1$ is available for the link $l$, then  $c$ is the first color of the path iff $x^{d,l}_{c-1} = 0$ and $x^{d,l}_{c} = 1$.
 If $(d, l, c-1) \not\in \lrvec U$, then $c$ is the first color of the path iff $x^{d,l}_{c} = 1$. 
 So, we identify that $c$ is the first color by checking that $x^{d,l}_{c-1} = 0$ and $x^{d,l}_{c} = 1$ in the case of $(d, l, c-1) \in \lrvec U$.
 The case $(d, l, c-1) \not\in \lrvec U$ is similar, so we provide the arguments only for the case $(d, l, c-1) \in \lrvec U$.

So, $\left(x^{d,l}_{c} - x_{d,l}^{c-1}\right) = 1$ if $x_{d,l}^{c-1} = 0$ and $x^{d,l}_{c} = 1$, and otherwise $\left(x^{d,l}_{c} - x^{d,l}_{c-1}\right) \leq 0$. In the latter case, the right side of Constraint~\eqref{mixedmilp:continious:spectrum} less or equals to $0$ and the constraint is vacuously true. 
If $x^{d,l}_{c-1} = 0$ and $x^{d,l}_{c} = 1$, then $\sum\limits_{\Delta c = 0 }^{w_d-1} x^{d,l}_{c+\Delta c} = w_d$ implies that $c$ is the first color of the path and all colors from $c$ to $c + w_d - 1$ are occupied by the path. Note that since Constraints (\ref{mixedmilp:div:s}-\ref{mixedmilp:div:s0}) implies that the value of flow equals $w_d$, no other colors can be occupied by the demand $d$.

In the case of $w_d > 1$, the flow can be split between several edges: it can happen so that for two links $l$ and $l'$ adjacent to $s_d$ the flow can occupy color $c$ of the link $l$ and color $c'$ of the link $l'$ (and may be other colors of these or other links). Constraints (\ref{mixedmilp:continious:spectrum}-\ref{mixedmilp:continious:spectrum2}) are not enough to solve this issue since they are defined only for the colors from the sets $C_{dl}$, so we also need Constraint~\eqref{mixedmilp:continious:spectrum3} that works for colors $c \not\in C_{dl}$ for which $(d, l, c) \in \lrvec U$.
If $c \not\in C_{dl}$, the color $c$ cannot be the first color of the path, but the color $c$ of the link $l$ still can be a part of some path for $d$ with the first color $c' \in C_{dl}$, such that $c' < c < c' + w_d$. So the color $c$ is occupied by $d$ if and only if the color $c-1$ is. Constraint~\eqref{mixedmilp:continious:spectrum3} ensures this condition.

Constraints (\ref{mixedmilp:continious:spectrum}-\ref{mixedmilp:continious:spectrum3}) imply that the flow cannot be split into multiple edges: Constraint~\eqref{mixedmilp:continious:spectrum3} ensures that if $c \not\in C_{dl}$ and $x^{d,l}_{c} = 1$, then $x^{d,l}_{c-1} = 1$ and there exists a color $c' < c$ such that $c' \in C_{dl}$, so the consecutive application of Constraint~\eqref{mixedmilp:continious:spectrum3} (to $c, c-1$, down to some $c'+1$) implies that Constraint~\eqref{mixedmilp:continious:spectrum} or \eqref{mixedmilp:continious:spectrum2} holds and ensures that the path occupies the contiguous range of colors of width $w_d$ of the link $l$. Constraint~\eqref{mixedmilp:div} guarantees that all these colors will be used on the next links in the path, and since the value of the flow is fixed and equals to $w_d$ according to Constraints (\ref{mixedmilp:div:s}-\ref{mixedmilp:div:s0}) the flow cannot be split: if $c$ is the first color and after the link $l$ goes link $l'$, then Constraints (\ref{mixedmilp:continious:spectrum}-\ref{mixedmilp:continious:spectrum2}) ensures that all the flow goes to $l'$, and Constraints (\ref{mixedmilp:div:s}-\ref{mixedmilp:div:s0}) ensures that there is no room for the other flow. Note
that if the flow leaves a vertex $u$, it leaves it by the only link provided that the path has no cycles, and this condition is ensured by the objective~\eqref{mixedmilp:objective}.  

\section{MILP model for the maximum subset problem}

We modify the model (\ref{mixedmilp:objective}-\ref{mixedmilp:continious:spectrum3}) for the maximum subset problem. 
Recall that the problem is to find a largest subset $D' \subseteq D$ such that the generalized RSA problem for $D'$ is solvable 
(and solve this problem). We add to the model Boolean variables $y_d, d \in D$ such that $y_d = 1$ means that the demand $d$ 
belongs to $D'$. Below we list only modified constraints, the rest of the constraints are the same.

\begin{align*}
    & \left(\sum_{ (d, l, c) \in \lrvec U} x^{d,l}_{c} - |U|\sum_{d\in D} y_d \right)  \to \min && &&  (\ref{mixedmilp:objective}')\\ 
    &\sum_{\substack{l \in N_+(s_d) \\ (d, l, c) \in \lrvec U}} x^{d,l}_{c} = w_d\cdot y_d & & \forall d\in D && (\ref{mixedmilp:div:s}')
\end{align*}

The value of the sum $\sum x^{d,l}_{c}$ is bounded by $|U|$ since 
variables $x^{d,\rvec l}_{c}$ and $x^{d,\lvec l}_{c}$ cannot simultaneously be equal to $1$ due to Constraint~\eqref{mixedmilp:unicolor}.
So the objective~(\ref{mixedmilp:objective}$'$) hits minimum values only if the sum $\sum_{d\in D} y_d$ hits its maximum.
Constraint~(\ref{mixedmilp:div:s}$'$) becomes inactive for $y_d = 0$, and works in the same way
as Constraint~\eqref{mixedmilp:div:s} for $y_d = 1$. Therefore, the modified model solves the maximum subset problem.
Since we add only $|D|$ new variables, while the model size $O(|V| + |U|)$ and $|U| \geq |D|$ in the nontrivial cases (otherwise the model is unsolvable), the modified model asymptotically the same size as the original model.

\section{Computational results}

We tested our MILP models using public open network topologies NSFNET and USNET. 
These networks were also used for experiments in~\cite{Xiong18}. 
The detailed computational results are uploaded to a Github repository~\cite{ExperimentsGit}.

\subsection{Test Case Generation}

To obtain non-trivial test-cases we routed paths in a network using shared path protection as follows. We randomly order all pairs of nodes and for each pair $(s,t), s \neq t$ we try to find a valid path $P$ that connects $s$ and $t$ and another valid path $R$ that does not share links with $P$. We call $P$ a \emph{main path} and $R$ a \emph{recovery path}. If such paths exist, we route them in the network. If two main paths are disjoint, then their recovery paths are allowed to share links, (hence, the name `shared' path protection). We repeat the process of path routing until both $P$ and $R$ exist for some pair of nodes $(s,t)$. When the process is finished, we remove all recovery paths from the network. This is how we generate the first kind of test cases.

It is easy to see that in the test cases of the first kind the answer to recovery problem is always positive, since each broken demand has a corresponding recovery path that does not intersect with the main path and no two recovery paths intersect if the corresponding main paths intersect. To obtain test cases of the second kind, we break a link in the test cases of the first kind and generate a new input network (without the broken link) by routing a path per every broken demand (the rest paths remained the same). This new networks are considered as input the same way as the network of the first kind, but in this case there are no guarantees that restoration is possible.


The value of the reachability threshold~$r_d$  depends on the \emph{Modulation Scheme}: 

\begin{itemize}    
    \item 5000 km for BPSK
    \item 2500 km for QPSK
    \item 1250 km for 8-QAM     
\end{itemize}

We considered the modulation schemes from the paper~\cite{Xiong18}, which also mentions the modulation scheme 16-QAM with the bound 625 km. We do not consider this scheme since this bound is too short for our needs. 

So a test case of the first kind is described by the network and modulation scheme, and a test case of the second kind also depends on the broken link. In each test case we use the same modulation scheme for each demand, but our method allows different modulation schemes for different demands as well. 
We consider the following test cases of the first kind:

\begin{itemize}
    \item \textbf{USNET}: BPSK, QPSK, 8-QAM
    \item \textbf{NSFNET}: BPSK, QPSK
\end{itemize}

%

To generate a congested routing with many demands we firstly routed demands of width 1 and then demands for width 4, 2, and 1 as well. So all colors of some links can be occupied by the paths of width 1, and we do not consider the test cases where such links are broken since it is important for our method to have demands of width greater than 1. So, we consider a link break scenario only for links that have at least one demand of width greater than 1. Because of that and the small reachability threshold for 8-QAM modulation, we have only 2 test cases for the routing scheme for USNET-8-QAM. The number of colors in all networks is the same for each link and equals 80.

\subsection{Tested Models}

Firstly, we tested simplified MILP models for the test cases of all demands with $w_d = 1$ without trimming. In this case the computations via noncommercial MILP solvers could take several hours (namely, CBC, SCIP, and GLPK). We did not add these results to repository\cite{ExperimentsGit}, since they have been tested on artificial networks of larger size.
In the case of all demands with $w_d = 1$  we got rid of Constraints~(\ref{mixedmilp:continious:spectrum}-\ref{mixedmilp:continious:spectrum3}), 
since in this case these constraints are vacuously true. 

The number of variables in the classical model is $\Theta(|L|\cdot|D|\cdot C)$ 
and the trimming significantly decreases the number of variables down to $1\%$ of the initial number.
So the presented model~(\ref{mixedmilp:objective}-\ref{mixedmilp:continious:spectrum3}) is solvable quite fast (in a few seconds)
even via noncommercial solvers (CBC and SCIP). 
Even hard test cases in our dataset had not been solved via the original model (without trimming) in 500 seconds, while most of the cases had been solved by the original model in less than 100 seconds.


In our tests we consider three models:

\begin{itemize}     
    \item \emph{Base}---the initial model (without trimming) 
    defined by Constraints~(\ref{mixedmilp:objective}-\ref{mixedmilp:continious:spectrum}) and Constraint~\eqref{mixedmilp:continious:spectrum2} that is used only for the case $c=0$; we set $x^{d,l}_c$ to $0$ if the color $c$ of the link $l$ is already occupied. 
    \item \emph{NoTrim}---the model defined by Constraints~(\ref{mixedmilp:objective}-\ref{mixedmilp:continious:spectrum3}) where we got rid of variables $x^{d,l}_c$  for which the color $c$ of the link $l$ is already occupied.
    \item \emph{Trimmed}---the model defined by Constraints~(\ref{mixedmilp:objective}-\ref{mixedmilp:continious:spectrum3}), where we got rid of all variables $x^{d,l}_c$ with useless triples $(d,l,c)$ found via trimming.
\end{itemize}

\subsection{Results}

In the tables below we refer to the tests of first kind as Net-Modulation and to the tests of the second kind as Net-Modulation $L$, where $L$ is the id of the broken link. The dataset and more detailed results are available in Github repository~\cite{ExperimentsGit}. For a few edges, solver time can exceed more than 100 seconds (for the Basic and NoTrim models). The solver's time limit is set to 500 seconds. Thus, if the computation exceeds 500 seconds, it is not completed. Note that it affects the mean time in tables below.
We also provide the aggregated time metrics with excluded hard test cases (that take over 100 seconds) in the repository~\cite{ExperimentsGit}. 

\begin{table*}[h]
    \vspace*{0.3in}
    \centering
    \begin{tabular}{|l|r||r|r|r||r|r|r||r|r|r|}
        \hline
         \multicolumn{2}{|c||}{} & \multicolumn{3}{c||}{Number of Variables} & \multicolumn{3}{c||}{Runtime of CBC Solver} & \multicolumn{3}{c|}{Runtime of SCIP Solver} \\ 
        \hline
                             \textbf{Net-Modulation }  & \#\textbf{Tests} & \textbf{Basic} & \textbf{NoTrim} & \textbf{Trim}
                                  & \textbf{Basic} & \textbf{NoTrim} & \textbf{Trim}
                                  & \textbf{Basic} & \textbf{NoTrim} & \textbf{Trim}
                               \\ \hline
        USNET-BPSK             &      41 & 146000 &  76600 &  9020    &  15.6  &   9.92 & \bf 1.69   &  39.23 &  34.43 &      2.72\\ \hline
        USNET-BPSK 23          &      40 & 148000 &  75700 &  7100    &  48.77 &  33.29 & \bf 0.78   &  51.36 &  47.23 &      1.36\\ \hline
        USNET-QPSK             &      37 &  99500 &  63800 &  2800    &  14.48 &  11.07 &     0.83   &  28.23 &  26.27 &  \bf 0.75\\ \hline
        USNET-QPSK 23          &      36 &  97300 &  62400 &  2700    &  19.6  &  11.61 &     0.84   &  31.07 &  28.53 &  \bf 0.68\\ \hline
        USNET-8-QAM            &       2 & 228000 & 218000 & 13600    & 670.17 & 526.72 & \bf 1.96   & 592.51 & 658.69 &      17.85\\ \hline
        USNET-8-QAM 31         &       3 & 214000 & 202000 &  5440    & 259.8  & 193.16 & \bf 1.11   & 566.57 & 361.64 &      9.20\\ \hline
        NSFNET-BPSK            &      21 &  64000 &  34900 &  6970    &   6.95 &   4.81 & \bf 1.07   & 116.06 &  32.29 &      1.77\\ \hline
        NSFNET-BPSK 5          &      20 &  67700 &  34400 &  4120    &  16.4  &  19.34 &     0.54   &  13.64 &  11.24 &  \bf 0.44\\ \hline
        NSFNET-QPSK            &      12 &  42600 &  32500 &  2300    &  25.54 &  61.85 &     0.61   &  47.47 &  26.49 &  \bf 0.44\\ \hline
        NSFNET-QPSK 19         &      11 &  46500 &  34700 &  1790    &  14.84 &  51.92 &     0.41   &  29.46 &  25.59 &  \bf 0.33\\ \hline
    \end{tabular}
    \smallskip
        \caption{All runtimes are in seconds on Apple M1 laptop.}\label{tab:mean_vars}
    \vspace{-2em}
\end{table*}

\if01

\begin{table}[h]
    \centering
    \begin{tabular}{|l|r|r|r|r|}
        \hline
                               & \#\textbf{Tests} & \textbf{Basic}  &  \textbf{NoTrim} & \textbf{Trimmed} \\ \hline
        USNET-BPSK           &      41 &  15.6  &   9.92 &   1.69 \\ \hline
        USNET-BPSK 23        &      40 &  48.77 &  33.29 &   0.78 \\ \hline
        USNET-QPSK           &      37 &  14.48 &  11.07 &   0.83 \\ \hline
        USNET-QPSK 23        &      36 &  19.6  &  11.61 &   0.84 \\ \hline
        USNET-8-QAM          &       2 & 670.17 & 526.72 &   1.96 \\ \hline
        USNET-8-QAM 31       &       3 & 259.8  & 193.16 &   1.11 \\ \hline
        NSFNET-BPSK          &      21 &   6.95 &   4.81 &   1.07 \\ \hline
        NSFNET-BPSK 5        &      20 &  16.4  &  19.34 &   0.54 \\ \hline
        NSFNET-QPSK          &      12 &  25.54 &  61.85 &   0.61 \\ \hline
        NSFNET-QPSK 19       &      11 &  14.84 &  51.92 &   0.41 \\ \hline
    \end{tabular}
    \smallskip
    \caption{Mean Time of Computations via CBC Solver}
    \label{tab:mean_time_CBC}    
\end{table}

\begin{table}[h]
    \centering
    \begin{tabular}{|l|r|r|r|r|}
        \hline
                               & \#\textbf{Tests} & \textbf{Basic}  &  \textbf{NoTrim} & \textbf{Trimmed} \\ \hline
        USNET-BPSK           &      41 &  39.23 &  34.43 &   2.72 \\ \hline
        USNET-BPSK 23        &      40 &  51.36 &  47.23 &   1.36 \\ \hline
        USNET-QPSK           &      37 &  28.23 &  26.27 &   0.75 \\ \hline
        USNET-QPSK 23        &      36 &  31.07 &  28.53 &   0.68 \\ \hline
        USNET-8-QAM          &       2 & 592.51 & 658.69 &  17.85 \\ \hline
        USNET-8-QAM 31       &       3 & 566.57 & 361.64 &   9.20 \\ \hline
        NSFNET-BPSK          &      21 & 116.06 &  32.29 &   1.77 \\ \hline
        NSFNET-BPSK 5        &      20 &  13.64 &  11.24 &   0.44 \\ \hline
        NSFNET-QPSK          &      12 &  47.47 &  26.49 &   0.44 \\ \hline
        NSFNET-QPSK 19       &      11 &  29.46 &  25.59 &   0.33 \\ \hline
    \end{tabular}
    \smallskip
    \caption{Mean Time of Computations via SCIP Solver}
    \label{tab:mean_time_SCIP}
\end{table}
\fi

The results show that our trimming technique significantly improves the model size and the runtime of the solver. Most test cases are solved using the CBC-solver in less than one second, and the hardest CBC test case takes a few seconds. The SCIP solver is slower on average, but it solves some test cases that are hard for the CBC-solver better in case of the Basic and NoTrim models. All in all, the Trimmed model has significantly less variables and can be solved by noncommercial solvers a hundred times faster than models without trimming. The NoTrim model is faster than the Basic model in many cases, but there are also test cases in which solvers solve the Basic model is faster and some of them are hard. So by the analysis of the NoTrim model we proved that the main impact of our model is based on the variables excluded via the trimming procedure, but not excluded because the corresponding colors have been occupied. Note that in some cases the infeasibility of the model can be proven during the trimming process and we met such cases in our dataset.

\section{Comparison with other models}

There are several known MILP approaches for (the original) RSA  problem, see~\cite{Bertero2018integer} for details. The difference of these approaches is of variable choice. Above we used the model with variables indexed by the demand, the link and the color, we call it demand-link-color indexing for short and call use similar naming for other approaches. There are other known models that use variables based on demand-color indexing~\cite{Velasco13} and demand-range indexing~\cite{Velasco2012}, where range means that for each demand there are variables that set the bound of occupied color's range, these models contain variables indexed by demands and links as well. More models based on other indexing were presented in~\cite{Bertero2018integer} including the demand-link-color indexing model that we used. 

While different indexing leads to different amounts of variables, the models observed in~\cite{Bertero2018integer} have the following bounds for the constraints: $\Theta(|D|^2|L|C)$, $\Theta(|D||L|C)$, and $\Theta(|D|^2|L|)$. So we choose our indexing, because we reduce via trimming not only the number of variables from $2|D||L|C$ down to $2|U|$, but the number of constraints as well: we have not $\Theta(|D||L|C)$ constraints but $O(|D||V|C+|U|)$ constraints. In fact even the number of Constraints~\eqref{mixedmilp:div} can be decreased since in practice there are typically cases where some demand $d$ does not have any path for several colors, so we remove constraints for nodes from a graph $G_c$ if $c$ is unavailable for~$d$.

The paper~\cite{Bertero2018integer} seems to be the first to discover the demand-link-color indexing-based model, in~\cite{Hadhbi19} presented another model. Since it is developed via a standard MILP technique and the paper had been published quite recently, 
suggesting that the MILP approach is not yet wide-spread in the communities interested in the RSA problem. However, our numerical results suggests that this is a useful method. 

Moreover, before recent development of the mentioned models, only models based on (in-advance) path generation had been spread widely. Such models have the disadvantage of exponential growth with the grow of the network's size, since the number of paths can be blown-up exponentially~\cite{Hadhbi19}. 
It looks like the main obstacle in development of MILP models that do not depend on generated paths is the constraint on contiguity of colors range. The rest of conditions are easily to construct from the well-known MILP-based approach for the flows: Constraints~(\ref{mixedmilp:div}-\ref{mixedmilp:div:s0}) define flows in independent networks corresponding to a demand and a graph $G_c$; Constraint~\eqref{mixedmilp:reachability} on reachability is quite obvious and so as Constraint~\eqref{mixedmilp:unicolor} that forbids occupation of the same edge (of the flow network) by different demands. 

Different indexing-based models uses different approaches to satisfy the contiguity constraint. The demand-link-color indexing-based model uses the standard MILP technique of constraint activation. Namely, a Boolean variable is set to 1 if the interested linear condition holds, and then we use this variable to activate another constraint, which vacuously holds if the variable is set to 0. We used the simplest direct implementation of this approach in the model for the maximum subset problem, in the case of contiguity Constraint~\eqref{mixedmilp:continious:spectrum} it looks a little bit more tricky since the Boolean condition 
\begin{equation}\label{eq:BoolCond}
    x^{d,l}_{c} \land \neg x_{c-1}^{d, l}
\end{equation}
can be expressed as
\begin{equation}
    b \geq x^{d,l}_{c} - x_{c-1}^{d, l}
\end{equation}
and the activation variable $b$ is triggered to equals $1$ only if assertion~\eqref{eq:BoolCond} holds. We used this approach in Condition~\eqref{mixedmilp:continious:spectrum2} to deal with gaps of variables (index) range that occurs because of the trimming and also use similar activation-based approach in Condition~\eqref{mixedmilp:continious:spectrum3} to completely fix the contiguity issue, since it became more complicated because of the trimming.

All in all, we used standard MILP techniques that were also used to construct a model for the RSA problem, but it looks like that these techniques are not so wide spreaded in the RSA-community. Our contribution is in the spreading this technique as well, but moreover we have combined MILP-based approach with a combinatorial approach (trimming) that allows significantly decreasing the model's size (in practice).

\section{Conclusion}

We adapted a known MILP model for the RSA problem to solve the ground truth problem for link restoration and the selection of maximum subsets of restorable demands. 
For this, we applied
\begin{itemize}  
    \item  combinatorial approaches to significantly reduced the number of MILP variables,
    \item appropriate choice of the model for a practical problem: restoration and associated maximum subset problem,
\end{itemize}

\bibliographystyle{plain}
\bibliography{refs}

\end{document}